\documentclass[12pt,preprint]{aastex}

\usepackage{emulateapj5}
\usepackage{apjfonts}

\newcommand{\euma}{\mbox{$\epsilon$~UMa}}

\shorttitle{Upper limit on oscillations in \euma}
\shortauthors{Retter et al.}

\begin{document}

\title {A Tight Upper Limit on Oscillations in the Ap star $\epsilon$ Ursae
 Majoris from {\em WIRE\/} Photometry}


\author{Alon Retter$^{1}$, Timothy R. Bedding$^{1}$, Derek L. Buzasi$^{2}$,
Hans Kjeldsen$^{3}$ and L\'aszl\'o L. Kiss$^{1}$}

\vskip 0.4 cm

\altaffiltext{1}{School of Physics, University of Sydney, 2006, Australia;
retter,bedding,laszlo@physics.usyd.edu.au}
\altaffiltext{2}{Department of Physics, 2354 Fairchild Drive, US Air Force Academy, 
CO 80840, USA; Derek.Buzasi@usafa.af.mil}
\altaffiltext{3}{Theoretical Astrophysics Center, University of Aarhus, 8000 
Aarhus C, Denmark; hans@ifa.au.dk}

\begin{abstract}

Observations of $\euma$ obtained with the star tracker on the Wide Field 
Infrared Explorer ({\em WIRE}) satellite during a month in mid-2000 are 
analyzed. This is one of the most precise photometry of an Ap star. The 
amplitude spectrum is used to set an upper limit of 75 parts per million 
for the amplitude of stellar pulsations in this star unless it accidentally 
oscillates with a single mode at the satellite orbit, its harmonics or 
their one day aliases. This is the tightest limit put on the amplitude 
of oscillations in an Ap star. As the rotation period of $\euma$ is 
relatively short (5.1~d), it cannot be argued that the observations were
made at a wrong rotational phase. Our results thus support the idea 
that some Ap stars do not pulsate at all.



\end{abstract}


\section{Introduction}

The excitation of pulsations in the lower part of the classical instability 
strip is still far from being understood. Some normal stars in this 
category show $\delta$~Scuti pulsations while others do not, and 
those that do pulsate only do so in a seemingly random subset of 
possible modes. Among the chemically peculiar stars, the so-called 
Ap stars that are the subject of this Letter, there is a similar 
puzzling division between the rapidly oscillating Ap stars (roAp 
stars) and the non-oscillating variety (noAp stars). This has lead 
some to propose that oscillations are present in noAp stars but at 
an undetectable level. Here we report one of the most precise 
photometric observations to date on an Ap star, which provides a 
strong upper limit on oscillations.

The Ap stars have peculiar spectra, with lines from elements such as 
Silicon, Magnesium, Mercury, Chromium, Europium and Strontium. The roAp 
stars have pulsations with periods about 5--15 min and typical amplitudes 
of a few milli-mag \citep{K1982, K1990, KM2000}. So far, thirty two 
objects have been classified in this subclass (Kurtz \& Martinez 2000;
Kurtz, personal communication). For recent theoretical works on the 
oblique pulsator model and the excitation mechanism in roAp stars see 
\citet{BPB2000}, \citet{CG2000}, \citet{BCD2001} and \citet{BD2002}.




A detailed study of the differences between roAp and noAp stars was 
carried out by \citet{HKM2000}. They found that as a group, the roAp 
stars are 0.57$\pm$0.34 mag brighter than the zero-age main sequence 
(ZAMS) while noAp stars are 1.20$\pm$0.65 mag above the ZAMS, which 
suggests that the latter are slightly more evolved than the roAp stars. 
It has also been proposed that an overabundance of the rare earth 
elements Nd and Pr may be a signature of roAp stars \citep{CB1998, 
WRK2000, RSM2001, RPK2002}. \citet{KLR2002} suggested, on the other 
hand, that all Ap stars in a certain temperature range pulsate and 
that the observed difference between roAp and noAp stars comes from 
the amplitude of the pulsation (i.e., it is simply below detectability 
in noAp stars). 


A key point in testing this last suggestion is to press down the 
observational upper limits on the pulsation amplitudes in noAp stars. 
\citet{KDC2003} observed the Ap star HD~965 over 2.7~h using a 2-m 
telescope and put an upper limit of 0.2 milli-mag  or $\sim$200 parts 
per million (ppm) on the amplitude of oscillations in the $B$-band.
This is the lowest limit for pulsation in a noAp star in the literature. 
However, the rotation period of HD~965 is uncertain and believed to be 
longer than two years. \citet{KDC2003} thus concluded that, although it 
is likely that HD~965 is a noAp star, their observations cannot reject 
the possibility that it is nevertheless a roAp star. The reason is that
according to the oblique rotator model, the amplitude of the oscillation 
is higher near maximum of the rotational modulation and may become zero 
at other phases \citep{K1990}. Therefore, \citet{KDC2003} stated that 
further observations of this star over a few years are required to 
substantiate its status as a noAp star. Here we report observations of 
the Ap star $\euma$, whose rotation period is short enough to avoid this 
problem.



$\euma$ (HR 4905 = HD 112185) is classified as an A0p Cr star 
\citep{BL1990}. Its effective temperature and radius were estimated 
as $\sim$9000K and $\sim4R_{\odot}$ \citep{PA1985,SB1979}. At V=1.8 
it is the brightest Ap star in the sky, but it is not known to pulsate.


A period of 5.0887 d was detected by \citet{G1931} in absorption line 
intensities of $\euma$ and was confirmed by further photometric and 
spectroscopic studies \citep{SH1943,P1953,WJ1980,WM1981}. \citet{P1953} 
found that the variation has a double wave structure with the two maxima 
being separated by half the period. The periodicity is understood as the 
rotation period of the star, but the strength of the magnetic field is 
still uncertain \citep{RW1990,DST1990,BL1990}.


\section{Observations and Reduction}

After the failure of the main mission of the Wide Field Infrared Explorer 
({\em WIRE}) satellite, launched by NASA in March 1999, its star tracker 
was successfully used for photometry of bright stars \citep{BCL2000, 
B2002, PBL2002, CAB2002, RBB2003}. 

The detector is a 512 $\times $512 SITe CCD with 27 $\mu$m pixels; each 
pixel corresponds to $\sim$1' on the sky. The camera is unfiltered and 
the effective response is roughly $V+R$. The observations of $\euma$ 
were obtained in the period 2000 June 22 -- July 21, with a cadence of 
0.5 seconds. Data reduction was done by applying a simple aperture 
photometry algorithm, which involved summing the central 4 $\times$ 4 
pixel region of the window of 8 $\times$ 8 pixels that can be read. The 
background level, due mainly to scattered light from the bright Earth, 
was estimated from the four corner pixels of each image, and subtracted. 
Following aperture photometry, any points deviating more than $2.5\sigma$ 
in mean flux, image centroid, or background level were rejected. The data 
points were then phased onto the satellite orbit, extreme points were 
rejected and the mean shape was subtracted. A final total of 5.7 million 
data points (out of about 6.1 million initial measurements) were obtained 
after applying these clipping procedures. Further details on the reduction 
method can be found in \citet{RBB2003}.



\section{Analysis}

The upper panel of Fig.~1 presents the smoothed light curve of $\euma$, 
where data from each satellite orbit have been binned into a single 
mean value, resulting in 308 points. Around JD 2451732 there are jumps 
of $\sim$10\% in the brightness of $\euma$. A comparison with the 
background level and the location of the centroid of the star on the 
CCD at this time (Fig.~1, three lower panels) shows that the effect is 
instrumental. The light curve is also affected by this effect, but to 
lesser extent, around JD 2451720 and 2451738. 

In an attempt to remove from the light curve the effect of minor 
shifts in the location of the star in the CCD and small changes 
in the background level, several decorrelation methods were tried 
\citep{BGN1991, RWB1995}. However, unlike for the case of our 
observations of Arcturus \citep{RBB2003}, these techniques did 
not satisfactorily correct the light curve. The reason for this 
behavior might be that, unlike Arcturus (which stayed within 
$\pm$0.03 pixel during the observing run), $\euma$ moved several 
tenths of a pixel from orbit to orbit. Consequently, the linear 
or low-order polynomial approximation of sub-pixel sensitivity 
variations was not valid over the range occupied by the star.


\begin{figure*}
\epsscale{0.8} 
\plotone{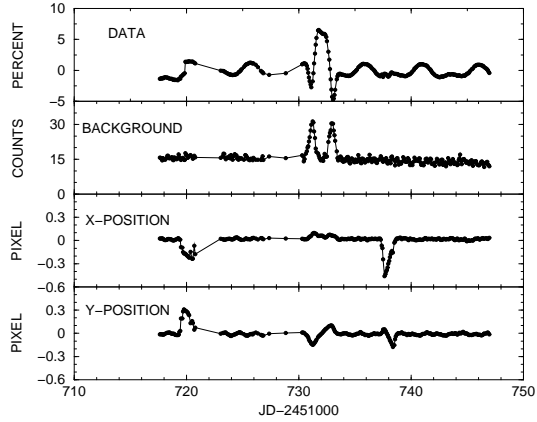}
\caption{Upper panel: The light curve of $\euma$ during the {\em WIRE} run 
in 2000 June-July. Each point represents a mean of about 18,600 0.1-s 
exposures obtained during a single 96-min orbit. 
Second panel: the background level. 
Third and bottom panels: the position of $\euma$ on the CCD in the X- and 
Y-axis (in pixels), respectively, relative to the mean location of the 
star in the run (258.21, 259.66).}
\end{figure*}




Fig.~2 displays the smoothed light curve of $\euma$ during the last two 
weeks of the observations, when the instrumental effects were minimal. 
The light curve clearly shows the presence of the 5.1-d rotation period 
(Section~1), which has a double-hump shape.


\begin{figure*}
\epsscale{0.8} 
\plotone{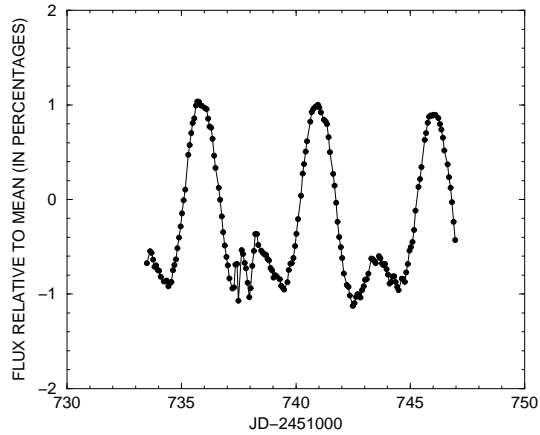}
\caption{The light curve of $\euma$ during the second half of the run, 
in which the instrumental effects have marginal influence.}
\end{figure*}



To search for oscillations (presumably around 10 min) in the data from 
the last two weeks of the run, the slow variation from rotation was first
subtracted. The upper panels of Figs.~3 \& 4 show the amplitude spectrum 
of this high-pass-filtered time series. The graph is dominated by strong 
peaks that correspond to the satellite orbit, its harmonics and, at lower
amplitudes, their 1-d$^{-1}$ aliases. The bottom panel in Fig.~4 shows 
the spectral window. The 1-d$^{-1}$ aliases are much stronger in the data
(relative to the satellite orbit and its harmonics) compared with the
spectral window. This behavior indicates that the data show instrumental 
variations within orbits and also with a periodicity of a day, perhaps
from variable reflection from Earth on the satellite. The second panels 
in the figures display the amplitude spectra after fitting and 
subtracting all these frequencies simultaneously. 



\begin{figure*}
\epsscale{0.8} 
\plotone{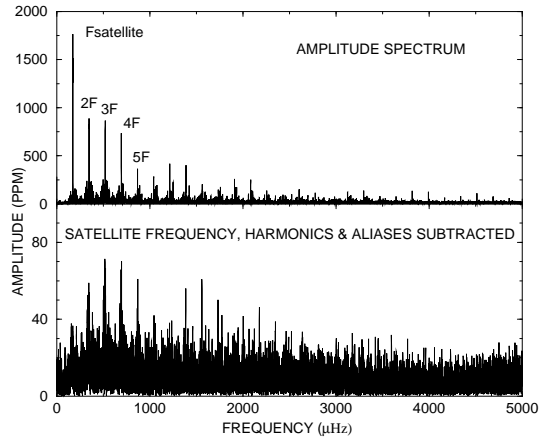}
\caption{Upper panel: The amplitude spectrum of $\euma$ during the last 
two weeks of the run, after the rotational variation was subtracted.
Fsatellite is the satellite orbital frequency (173.6 $\mu$Hz) and 2F, 
3F... represent its harmonics.
Bottom panel: The amplitude spectrum after subtracting variations at
the satellite orbital frequency, its harmonics and their 1-d$^{-1}$ 
aliases. Note the change in vertical scale.}
\end{figure*}


\begin{figure*}
\epsscale{0.8} 
\plotone{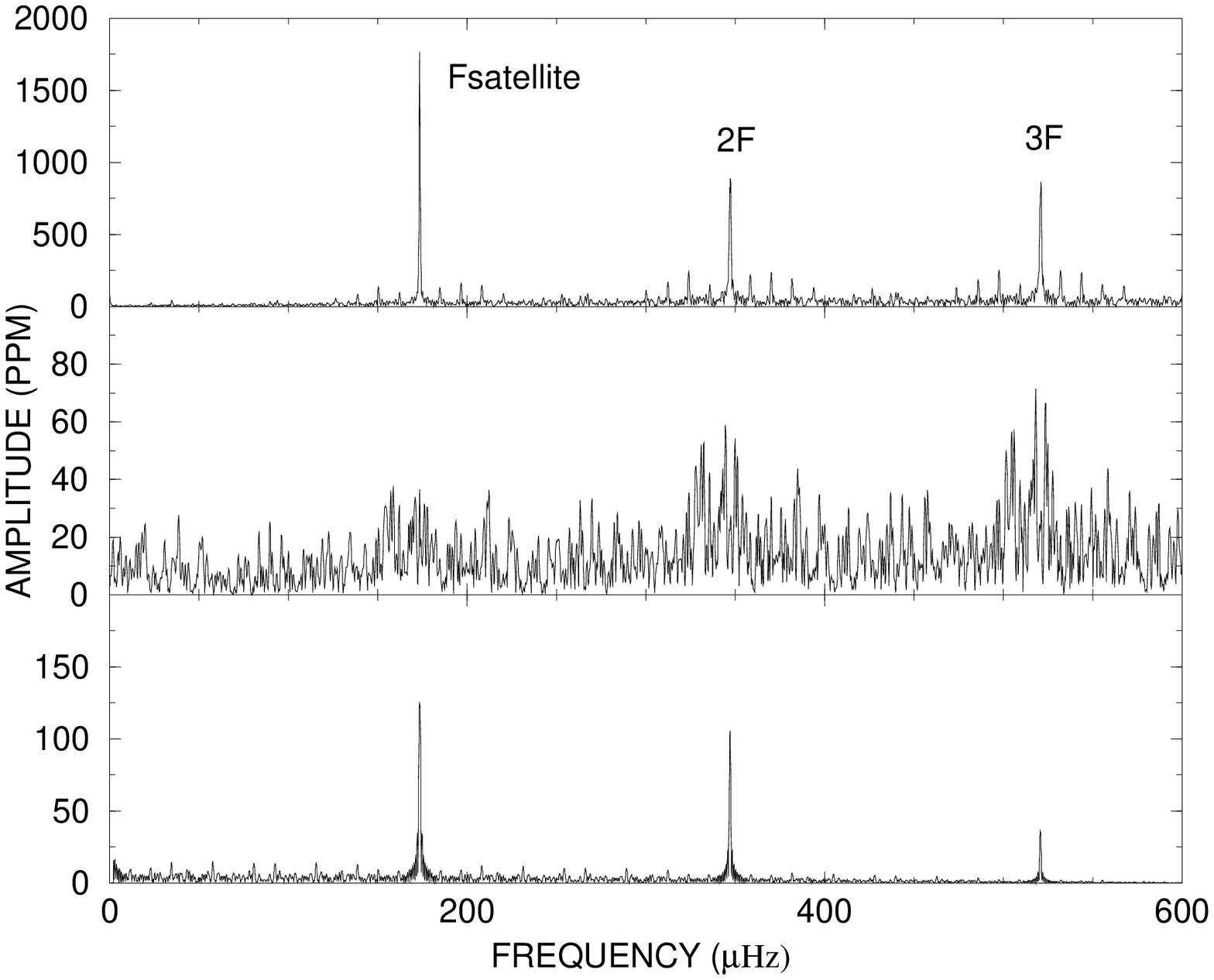}
\caption{Same as Fig.~3 for the lowest frequencies. The presence of the 
1-d$^{-1}$ aliases around both the satellite frequency and its harmonics 
is evident. The lower panel presents the spectral window}
\end{figure*}


All peaks in the residual amplitude spectrum (which are not harmonics or 
1-d$^{-1}$ aliases) have amplitudes below 75 ppm. Most of the highest 
remaining peaks are, however, close to the harmonics of the satellite 
orbit and all appear in the spectral window. We conclude that there are 
no oscillations with amplitudes above 75 ppm.  

The possibility cannot be ruled out that the star oscillates with a 
single mode that falls exactly on the satellite orbit, its harmonics 
or their 1-d$^{-1}$ aliases. We estimated the probability of this 
occurring by chance by dividing the typical width of the peaks 
(taken as the full-width at half maximum, which is set by the length 
of the observing run) by 16 d$^{-1}$ (the satellite orbit, for higher 
amplitudes) and by 1 d$^{-1}$ (for lower amplitudes, since the data 
suffer from 1-d$^{-1}$ aliases). The probabilities are about 0.5\% 
and 7\%, respectively. In other words there is a 0.5\% chance that it 
oscillates with an amplitude below $\sim$1800 ppm and a 7\% chance 
for an amplitude below $\sim$250 ppm.




To check whether the star has stronger pulsations at a specific phase of 
the 5.0887-d rotation period (see Section 1), the data from the last two 
weeks of the run (after removing the variation of the rotation period) 
were divided into ten equal phase bins. The corresponding power spectra of 
the bins were consistent with the power spectrum of the whole data (Figs.~3 
\& 4) and no significant peaks were found (besides the satellite orbital 
frequency, its harmonics and their 1-d$^{-1}$ aliases).

\section{Discussion}

The {\em WIRE} photometry clearly shows that $\euma$ is variable on 
time scales of a few days with a period of $\sim$5 days (Fig.~2). 
Our results confirm the previous detections of the rotation period 
(Section~1). The noise in our data is comparable to that in the 
highest precision photometry done so far on roAp stars obtained 
during several years \citep{KWR1997} or by the WET collaboration 
\citep{KKR2003}.


Of major interest is the question whether $\euma$ pulsates or not. We 
found an upper limit of 75 ppm on oscillations in this star, unless by 
chance it has a single mode that falls on the satellite orbit, its 
harmonics or their 1-d$^{-1}$ aliases. This, however, is very unlikely 
since roAp stars usually show several pulsation periods in their light 
curves \citep{K1990}.


We note that the observations were done using an effective response of 
roughly $V+R$ (Section~2). The amplitude of pulsations in roAp stars 
decreases with wavelength \citep{KM1996, MK1998, M2000}. $\euma$ may, 
therefore, pulsate with amplitudes larger than the limit given above in 
narrower and bluer bands. Using the detector response, a typical spectrum
of an Ap star \citep{LBP2003} and Fig. 1 of \citet{MK1998} we estimated 
that the limits we can put on the amplitude of pulsations in $\euma$ are 
about 50\% and 120\% larger in $V$ and $B$, respectively. \citet{KM1996}, 
\citet{MK1998} and \citet{M2000} found that the phase of the modulation 
also depends on the wavelength. This effect is, however, relatively 
small and the phase shift between the optical filters is about 0.5 
radians. Therefore, different colors are probably not anti-phased with 
each other and cannot cancel the effect of pulsations. Using simulations, 
we found that in our case the reduction of the amplitude by this effect 
is negligible (less than 3\%).




The rotation period of $\euma$ is relatively short at 5.1~d (Sections~1 
\& 3) and the {\em WIRE} observations covered several cycles of this 
periodicity. Therefore, it cannot be claimed that the non-detection of 
oscillations presented in this work is because of observations at a 
wrong rotation phase, which may be the case for noAp stars with long
rotation periods (Section 1). In fact, we could not find oscillations 
at specific phases (Section 3). Our results thus suggest that $\euma$ 
does not oscillate at all. It is interesting to note that it was 
proposed that an anomaly in the abundances of Nd and Pr may be a 
signature of roAp stars (Section~1). $\euma$ does not have this 
anomaly and the null result on oscillations in this star thus supports 
this suggestion.

It was suggested that all Ap stars in a certain temperature range 
pulsate \citep{KLR2002}. The temperature of $\euma$ was estimated as 
$\sim$9000K (Section 1), which is slightly higher than the range of 
roAp stars -- 7000--8300K \citep{K1990}, although these limits are 
somewhat uncertain. It is therefore possible that the non detection
of oscillations in $\euma$ is simply because the star is too warm 
for the instability strip for roAp stars.
 
The question of whether noAp stars do pulsate with amplitudes below the 
detection limit is still open. Future space missions that significantly 
decrease the upper limit on the amplitudes of pulsations in these stars 
may give an answer to this question.





\acknowledgments

We are grateful to all the people at NASA headquarters, GSFC and IPAC
who made the use of the {\em WIRE} satellite possible. We thank the 
anonymous referee for useful comments and Don Kurtz for helpful 
discussions. AR, TRB and LLK are supported by the Australian Research 
Council and HK by the Danish Natural Science Research Council, and the 
Danish National Research Foundation through its establishment of the 
Theoretical Astrophysics Center. DLB acknowledges support from NASA 
(NAG5-9318). 






\end{document}